\documentclass{iopart}

\usepackage{graphicx}

\begin{document}

\title{Variational Average-Atom in Quantum Plasmas (VAAQP) - first numerical results.}

\author{R Piron$^1$\footnote{Electronic address: robin.piron@cea.fr}, T Blenski$^1$ and B Cichocki$^2$}
\address{$^1$ CEA, IRAMIS, Service des Photons Atomes et Mol\'ecules, F-91191 Gif-sur-Yvette, France}
\address{$^2$ Institute of Theoretical Physics, Warsaw University, Ho\.{z}a 69, PL 00-681 Warsaw, Poland}


\date{\today}

\begin{abstract}
The work on a new fully variational model of average-atom in quantum plasmas using a numerical code called VAAQP is reported. A brief description of the code is given. Application to aluminium at solid density and temperatures between $0.05$ and $12$ $eV$ is presented. Comparisons to results obtained using other approaches are also shown and discussed. The results prove the feasibility of the variational model in the warm dense matter regime. Effects of the variational treatment can lead in this region to significant differences with respect to existing models.
\end{abstract}


\submitto{\JPA}

\section{Introduction}
In this paper, we present the first application in a numerical code (VAAQP) of a new fully variational model of quantum plasma at thermodynamical equilibrium. Models of screened ions in equilibrium plasmas with quantum all electrons are useful to photo-absorption and equation of state calculations. The existing models \cite{Liberman79,PerrotReport,Wilson06} are not fully variational in the sense that they do not respect the virial theorem. Recently a fully variational model respecting the virial theorem has been proposed \cite{Blenski07a,Blenski07b}. The variational character of the model gives a firm basis to the atomic structures calculations and provides electronic pressures that are thermodynamically coherent.

Among the most interesting applications are cases corresponding to the Warm Dense Matter (WDM) regime. Special attention is thus paid to matter at high density (ie. solid density) and relatively low temperature (around few $eV$ or below). The main problem in such regime comes from the long range Friedel oscillations.

This paper is organized as follows. Sec. II contains a short review of the model. Sec. III is devoted to a brief description of the VAAQP code. Some results obtained with the code for aluminium ($Al$) in the WDM regime are presented in Sec IV. Differences between models requiring the neutrality of the WS sphere and the variational model are discussed in this last section.

\section{Variational theory of average-atom in jellium}
The VAAQP code uses a new model \cite{Blenski07a,Blenski07b} of quantum plasma at thermodynamical equilibrium with bound and continuum electrons treated within the same formalism. In this model, the equilibrium is determined by 3 parameters: atomic number $Z$, temperature $T$ and ion density $n_i$. All other quantities, including mean ionization $Z^*$, are determined from variational equations using for instance the density functional theory (DFT).
In the Thomas-Fermi (TF) approximation to the electron density the present model leads naturally to the Thomas-Fermi ion-in-cell approach \cite{Feynman49}.

The starting point of the present model is the cluster expansion \cite{Felderhof82,Blenski92,Felderhof95a} from which the two first orders are retained:
\begin{equation}
\label{eq_cluster_f}
f\left(n_i,Z,T\right)=f_0+\langle f\rangle_1+...
\end{equation}
$f_0$ is the free energy per unit volume of an uniform electron gaz with an unknown density $n_0$,
\begin{equation}
\label{eq_def_Z}
n_0=n_i Z^*
\end{equation}
In the zeroth order, electron charge density is neutralized by an uniform ion background. $\langle f\rangle_1$ is the average contribution of one ion immersed in that jellium to the free energy density. In the first order, the system is no more homogeneous.
The cluster expansion leads to:
\begin{equation}
\langle f\rangle_1=n_i\int\,d^3r\left\{f_1^{ion+jellium}\left(X;n_i,Z,T;\vec{r}\right)-f_0\left(n_0,T\right)\right\}
\end{equation}
The substraction of the zero order term assures the convergence of the above integral.
All structure variables, including $n_0$ are denoted by $X$.

The free energies per ion are thus defined as follows:
\begin{equation}
F_0=F_0\left(n_0,n_i,T\right)\equiv\frac{f_0\left(n_0,T\right)}{n_i}
\end{equation}\begin{equation}
F\left(X;n_i,Z,T\right)=F_0+\Delta F_1
\end{equation}\begin{equation}
\Delta F_1=\int\,d^3r\left\{f_1^{ion+jellium}\left(X;n_i,Z,T;\vec{r}\right)-f_0\left(n_0,T\right)\right\}
\end{equation}
All structure variables $X$ are determined from the minimization of the free energy $F\left(X;n_i,Z,T\right)$ with additional conditions:
\begin{itemize}
\item The first order non-central-ion distribution is not uniform but has the form of a cavity with a radius $R$:
\begin{equation}
n_0 \theta(r-R)
\end{equation}
\item The neutrality of all charges, taking the cavity into account, is respected:
\begin{equation}
\label{eq_neutr_all_charges}
Z+\int\,d^3r\left\{n(\vec{r})-n_0 \theta(r-R)\right\}=0
\end{equation}

\item The total electron number per unit volume has the following cluster expansion:
\begin{equation}
\label{eq_clust_el_num}
Z n_i=n_0+n_i\int\,d^3r\left\{n(\vec{r})-n_0\theta(r-R)\right\}+...
\end{equation}
Using Eq.~\ref{eq_clust_el_num} in Eq.~\ref{eq_neutr_all_charges} leads to the identity: $R=R_{WS}$
\end{itemize}

The minimization leads to the SCF equations plus a new one which allows to determine $n_0$ or $Z^*$:
\begin{equation}
\label{eq_variational}
\int\,d^3r\left\{V_{el}(\vec{r})\theta(r-R)\right\}=0
\end{equation}
where $V_{el}(r)$ is the electrostatic part of the SCF potential \cite{Blenski07a,Blenski07b}.

The main difference between \textsc{Inferno}-type models \cite{Liberman79,Wilson06}  and the present one is the way of determining the mean ionization $Z^*$. In the \textsc{Inferno} type models, it stems from the neutrality of the WS sphere whereas in the present one it is obtained from the variational principle (\ref{eq_variational}). In this model, the WS sphere is in general non-neutral.

\section{The VAAQP code}
\begin{figure}[t]
\centerline{\includegraphics[width=8cm]{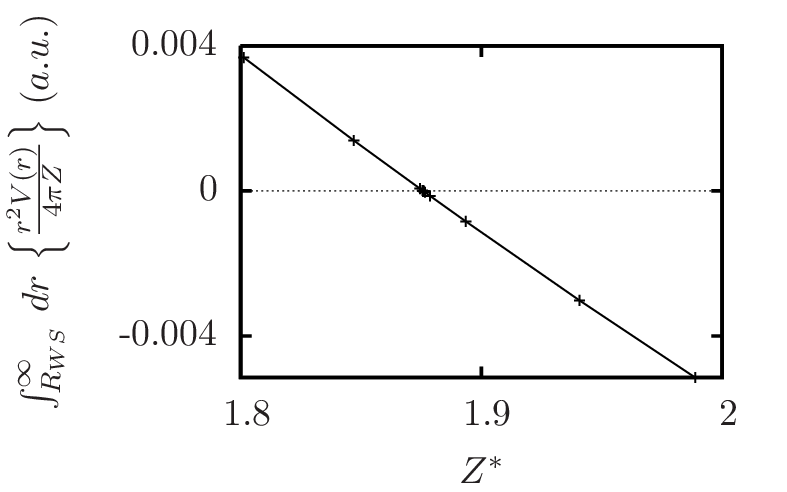}}
\caption{Left hand term of Eq.~\ref{eq_variational} versus $Z^*$. Example of solution in the case of $Al$ plasma at $5$ $eV$ temperature and $2.7$ $g.cm^{-3}$ matter density. Each point on the figure corresponds to a SCF solution in the four-parameter space ($Z$, $T$, $n_i$, $Z^*$).
\label{fig1}}
\end{figure}

The VAAQP code solves the equations of the variational model (see Eqs.~\ref{eq_cluster_f}-\ref{eq_variational}). First, solutions to the SCF equations in the four-parameter space ($Z$, $T$, $n_i$, $Z^*$) are calculated. Then among them is found such that fulfills Eq.~\ref{eq_variational} (see FIG.~\ref{fig1}).
The code has three options to calculate the electron density: semiclassical (Thomas-Fermi), quantum-non-relativistic (Schroedinger) or quantum-relativistic (Dirac) formalism.

\begin{figure}[b]
\centerline{\includegraphics[width=8cm]{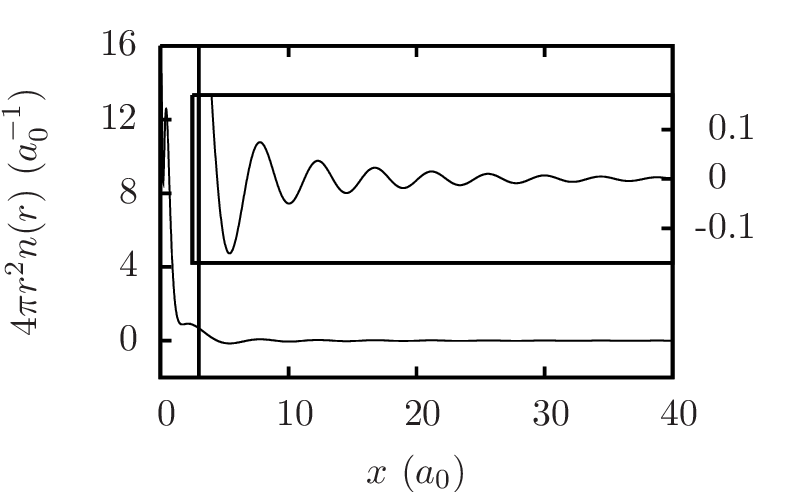}}
\caption{Plot of the electron density versus radius for $Al$ at $0.25$ $eV$ temperature and $2.7$ $g.cm^{-3}$ matter density with a detailled view of the Friedel oscillations.
\label{fig2}}
\end{figure}
In the case of quantum calculations, continuum states contributions to observables are integrated over energy using a resonnance-catching adaptative-mesh-refinement method.
Electron density is numerically computed for radius from zero up to an asymptotic value where it is matched to the asymptotic form given by the linear response theory of an homogeneous dense plasma \cite{Khanna76,Gouedard78}. Solutions are considered valid if they are unsensitive to choice of the matching radius and are not mesh-dependent. The matching radius depends mostly on the range of the Friedel oscillations and becomes larger as temperature is lower (see FIG.~\ref{fig2}).

Instead of looking for solution to Eq.~\ref{eq_variational}, the code can be set up to fulfill other conditions as for instance the neutrality of the WS sphere.

\section{Model comparison in case of warm dense aluminium}
The code has been applied in the case of $Al$ at typical condition of WDM ie. solid density ($2.7$ $g.cm^{-3}$) and temperatures between $0.05$ and $12$ $eV$. In this regime Friedel oscillations decay on a scale much larger than the WS radius (decay length $(2.b_0^F)^{-1}=55.2$ $a_0$ at $0.05$ $eV$ temperature, WS radius $R_{WS}=2.99$ $a_0$). In such cases, region of numerical calculations should extend over several tens of $R_{WS}$. It is important to recall that the present approach is based on the assumption that the SCF potential decays exponentially due to screening which is only valid at finite temperatures \cite{Khanna76,Gouedard78}.

The exchange-correlation term is taken in the local density approximation (LDA). For this exploratory study, in order to have direct comparisons with the Thomas-Fermi-Dirac (TFD) model, Dirac exchange term \cite{Dirac30,KohnSham65a} was used. Iyatomi and Ichimaru finite-temperature exchange-correlation term \cite{Iyetomi86} has also been tested. In the temperature regime of interest in this study, the differences between results obtained using these two approximations appeared to be relatively small.

\begin{figure}[t]
\centerline{\includegraphics[width=8cm]{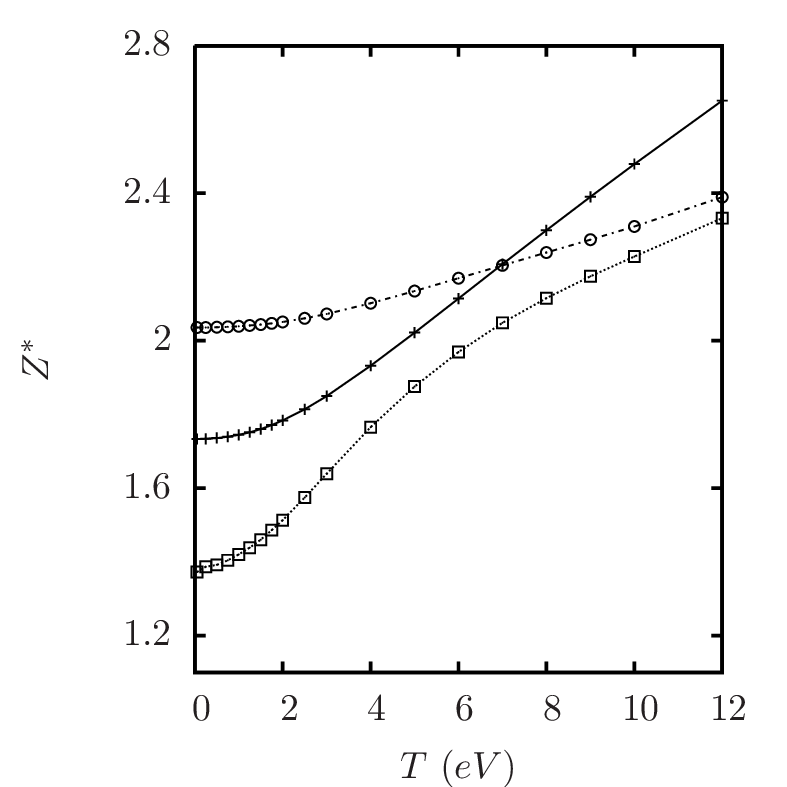}}
\caption{Mean ionization $Z^*$ as defined in Eq.~\ref{eq_def_Z} versus temperature in case of  $Al$ at $2.7$ $g.cm^{-3}$ matter density. Solid line corresponds to variationnal model with Thomas-Fermi density. Plus symbols denote neutral WS sphere model with Thomas-Fermi density. Squares corresponds to variationnal model with non-relativistic quantum density. Circles denote neutral WS sphere model with non-relativistic quantum density.
\label{fig3}}
\end{figure}

FIG.~\ref{fig3} displays the behavior of the mean ionization $Z^*$ that was defined in Eq.~\ref{eq_def_Z}. This definition is the only one justified in the framework of \cite{Blenski07b} and can be applied to any model as it corresponds to the asymptotic value $n_0$ of the electron density, which is related to the chemical potential. It is important to notice that the notion of unbound electrons is not related to any well defined observable of a quantum operator. In the quantum approaches other definitions are sometimes used. One of those is the value of the electron density at the WS radius. This definition is motivated by the fact that it is identical to the previous one in the Thomas-Fermi case. Another one is the difference between the atomic number $Z$ and the sum of the bound levels occupation numbers. These other definitions will not be considered in this paper.

\begin{figure}[t]
\centerline{\includegraphics[width=8cm]{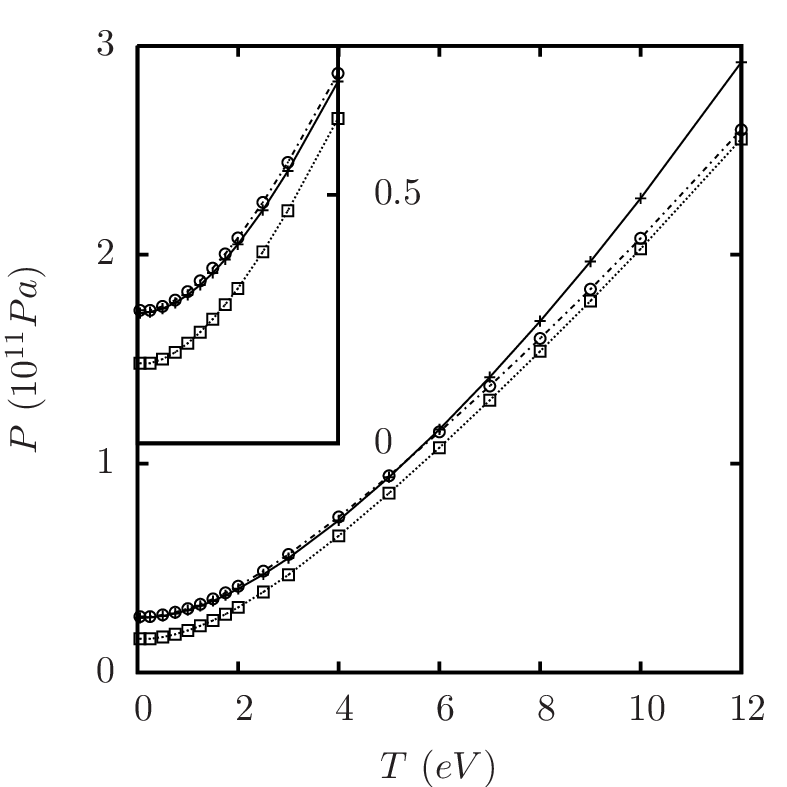}}
\caption{Electronic pressure $P$ versus temperature in case of  $Al$ at $2.7$ $g.cm^{-3}$ matter density. Solid line corresponds to variationnal model with Thomas-Fermi density. Plus symbols denote neutral WS sphere model with Thomas-Fermi density. Squares corresponds to variationnal model with non-relativistic quantum density. Circles denote neutral WS sphere model with non-relativistic quantum density.
\label{fig4}}
\end{figure}

Results of four approaches are presented on FIG.~\ref{fig3}. The first one is the result of the variational model with electron density calculated using the non-relativistic quantum formalism.  The second one is the result of the model with electron density calculated using the non-relativistic quantum formalism but in which the Eq.~\ref{eq_variational} was replaced by the condition of the neutrality of the WS sphere (NWS):
\begin{equation}
\label{eq_neutr_WS}
Z=\int_{|\vec{r}|\leq R_{WS}}d^3r\,n(\vec{r})
\end{equation}
The condition based on Eq.~\ref{eq_neutr_WS} is used in the \textsc{Inferno} model. However calculations reported here with Eq.~\ref{eq_neutr_WS} do not correspond exactly to the \textsc{Inferno} model as in our case the potential outside the WS sphere is not set to be zero.
Two calculations using the TF formalism are also reported, one with Eq.~\ref{eq_variational}, another with Eq.~\ref{eq_neutr_WS}. According to \cite{Blenski07a,Blenski07b}, in the TF case, these two approaches are stricly equivalent which is confirmed by FIG.~\ref{fig3} and FIG.~\ref{fig4}.

As it is seen on FIG.~\ref{fig3}, results from quantum calculations with Eq.~\ref{eq_variational} can differ significantly from those obtained using Eq.~\ref{eq_neutr_WS}. Differences are especially pronounced at low temperature and tend to vanish as temperature increases. In the case of $Al$ at solid density and at temperature below $2.5$ $eV$, the atomic structure is different in the two quantum models. The $3s$ shell is found to exist whithin the variational model whereas it is absent when using the NWS model. These differences are mainly related to the quantum behavior of the density outside the WS sphere, namely to the Friedel oscillations wich are present in this regime.

Electronic pressure is calculated from the formula:
\begin{equation}
\label{eq_press}
P=-f_0+n_0\left(\mu+V_{xc}(n_0)+V_{el}(R_{WS})\right)
\end{equation}
$f_0=f_0^0+f_0^{xc}$ is the free energy per unit volume of an uniform electron gaz including exchange-correlation. $f_0^0$ is the perfect gas free energy for the grand-canonical ensemble, namely in atomic units:
\begin{equation}
f_0^0=-\frac{2}{3}\frac{\sqrt{2}}{\pi^2}T^{{5}/{2}}I_{{3}/{2}}\left(\frac{\mu}{T}\right)+n_0\mu
\end{equation}
In the present study $f_0^{xc}$ was set to $\epsilon_{xc}$ from \cite{KohnSham65a}. $V_{xc}$ is the exchange-correlation part of the SCF potential \cite{Blenski07a,Blenski07b}.
\begin{equation}
V_{xc}(n)=\frac{\partial f_0^{xc}(n)}{\partial n}=-\left(\frac{3 n}{\pi}\right)^{1/3}
\end{equation}
Eq.~\ref{eq_press} is obtained from the variational theory and applies only to the case where Eq.~\ref{eq_variational} is fulfilled. The quantum NWS model is not variational and the use of the above pressure formula is only given for comparison purposes. In the case of models that are not fully variational the electronic pressure is often calculated by numerical differentiation of the free energy with respect to the ion density.

As can be seen on FIG.~\ref{fig4} relative differences in calculated pressures from the two quantum models can be about 40\% in the low temperature region and decrease as temperature increases. Pressures obtained using the quantum NWS model at low temperatures seem to be relatively close to those obtained from the TFD model.

FIG.~\ref{fig3} and FIG.~\ref{fig4} both display curves that are continuous with respect to temperature. In the case of the quantum variational model, it is worth to stress that continuity is preserved despite the fact that the $3s$ shell is disappearing between $2$ and $2.5$ $eV$. This is due to a careful treatment of continuum resonnance in the VAAQP code.

\section{Conclusion}
In this paper, a fully variational model of the average atom in quantum plasmas is applied for the first time in a numerical code called VAAQP. It is shown on some chosen examples that the code is able to calculate variational self-consistent equilibrium at high matter density and relatively low temperatures. In this regime, Friedel oscillations of the electron density and SCF potential present a real challenge from the numerical point of view.

Effects of the variational treatment is studied by comparison to results from a model requiring the neutrality of the WS sphere. The variational model can lead to pressures, mean ionizations and atomic structures that differ significantly from those obtained using existing models especially in the low temperature region.

The work on the present model is in progress. Among objectives is a coherent approach to the equation of state and the radiative properties of the warm dense matter.


\end{document}